
\input harvmac
\sequentialequations
\def\half{{\textstyle{1\over 2}}}  
\def\quart{{\textstyle{1\over 4}}} 
\def\3{\,{^3}\!}     
\def\s{\,{^*}\!}     
\def\V{\ V\!\!\!\!\!_{_\sim}\  }   
\def\epsi{\epsilon\!\!\!_{_\sim}\,}%
\def\CQG#1{Class. Quantum Grav. {\bf #1}}
\def\NP#1{Nucl. Phys. {\bf #1}}
\def\PL#1{Phys. Lett. {\bf #1}}
\def\PR#1{Phys. Rev. {\bf #1}}
\def\PRL#1{Phys. Rev. Lett. {\bf #1}}
\Title{DFUPG-83/94}{Regge calculus and Ashtekar variables}
\centerline{Giorgio Immirzi\footnote{$^\dagger$}
  {e--mail: immirzi@perugia.infn.it}}
\bigskip\centerline{Dipartimento di Fisica, Universita' di Perugia,}
\centerline{and Istituto Nazionale di Fisica Nucleare,
sezione di Perugia}
\centerline{via A. Pascoli, 06100 Perugia, Italy}

\vskip .3in
Spacetime discretized in simplexes, as proposed in the
pioneer work of Regge, is described in terms of
selfdual variables. In particular, we elucidate
the "kinematic" structure of the initial value problem, in
which 3--space is divided into flat tetrahedra, paying
particular attention to the role played by
the reality condition for the Ashtekar variables.
An attempt is made to write down the vector and scalar
constraints of the theory in a simple and potentially useful way.

\Date{2.2.94}
%
\newsec{Introduction.}
  Discussing the (1971) future of Regge calculus
\ref\regge{T. Regge, Il Nuovo Cim. {\bf 19} (1951) 558.},
Misner, Thorne and Wheeler
\ref\MTW{C.W. Misner, K.S. Thorne and J.A. Wheeler, "Gravitation",
Freeman, San Francisco, 1971.}
expressed the hope that "Regge's truly geometric way of formulating
general relativity will someday make the content of the Einstein
field equations ... stand out sharp and clear,...". Perhaps
this hope has not been fulfilled (though not for want of trying),
but one may still turn to Regge calculus for an intuitive
geometric interpretation of formal developments, or for a clean and
geometrically meaningful regularization of the theory.

In the latest and very promising attempt to quantize gravity,
this time in loop space
 \ref\carlolee{C. Rovelli and L. Smolin, \PRL{61} (1988) 1155
  \semi C. Rovelli, \CQG{8} (1991) 1613.},
it is the lack of a reliable regularization that hampers
the progress towards a consistent theory.
A regularization is needed to deduce the form of the constraints,
and  inevitably this breaks the backround independence that is
central to the whole program. Worse still, when the cutoff is
removed, backround independence is not recovered: the action of
the Hamiltonian constraint on the wave function can be expressed
in terms of genuine loop operations, weighted with the angles
between the branches of the loop at an intersection
\ref\miles{M. P. Blencowe, \NP{B341} (1990) 213
 \semi R. Gambini, \PL{B255} (1991) 180
 \semi B. Br\"ugmann and J. Pullin, \NP{B390} (1993) 399.}.
With such an obscure expression it is impossible to
decide whether the constraint algebra closes or not, quite apart
from the technical difficulty of the calculation.

This situation could perhaps be remedied if one could describe
Regge's discretized space in terms of the
new canonical variables for general relativity introduced
by A.Ashtekar \ref\abhaya{A. Ashtekar, \PRL{57} (1986) 2244;
\PR{D36} (1986) 1587.}
\ref\abhayb{A. Ashtekar, "Non--perturbative canonical quantum
gravity", notes prepared in collaboration with R. Tate,
  World scientific, Singapore, 1991.},
on which the loop representation is based.
This program was begun by V. Khatsymovsky
\ref\khats{V. Khatsymovsky, \CQG{6} (1989) L249.},
and there have been other attempts  to formulate a discretized
canonical theory along these lines
\ref\lee{P. Renteln and L. Smolin,  \CQG{6} (1989) 275
 \semi M. Miller and L. Smolin, Syracuse preprint, gr-qc 9304005.}.
In this work we complete the "kinematical" part of such a program,
but do not attempt to formulate a canonical theory, nor to
investigate in detail the structure of the constraints.

Quite apart from the aims stated above, a formulation of this type
might have an interest of its own because it is to some
extent modeled on recent work by G. 't Hooft
\ref\thooft{G. 't Hooft,  \CQG{10} (1993) 1653.}
 in $2+1$ dimensions.
It appears that, in this admittedly simpler case, on can go
surprisingly far in the construction of a quantum theory, in
particular on the crucial problem of the
connection with the Euclidean formulation,  on the formulation
of a canonical theory and on the analysis of the constraints
\ref\wael{H. Waelbroeck,  \CQG{7} (1990) 751.}.

We recall in \S 2  the basic ideas of Regge calculus, in the
tetrad form developed by various authors
\ref\bander{M. Bander, \PRL{57} (1986) 1825,
    \PR{D36} (1987) 2297, {\bf D38}  (1988)
 \semi V. Khatsymovsky, \CQG{8} (1991) 1205
 \semi H. C. Ren, \NP{B301}  (1988) 661
 \semi M. Caselle, A. D'Adda and L. Magnea, \PRL{232}  (1989) 457.}
in the last
few years, and their translation in terms of selfdual variables.

In \S 3 we analyze the initial value picture one obtains
separating 3--space from time, emphasizing in particular the
definition and the properties of the Ashtekar variables.
In a 3--manifold divided in tetrahedra, there is a set of variables
$\tilde E^{a\mu}$ for each tetrahedron,
while the transport matrices $L(AB)_{ab}$, related to the
Ashtekar connection $A^a_\mu$, are associated to the links of the
dual lattice. We use the reality conditions
\ref\ART{A. Ashtekar, J.D. Romano and R.S. Tate,
\PR{D40} (1989) 2572.} \abhayb , to
derive the form of the curvature one finds associated to
the polygons in the dual lattice, and find that it is consistent
with one would expect from the continuum case
\ref\me{G. Immirzi, \CQG{10} (1993) 2347.}.

We attempt in \S 4 to express the scalar and the vector constraints
of the Ashtekar formulation in a way that is suitable to be
applied to Regge calculus, using  arguments based on the weak
coupling case and on an intuitive picture borrowed from
\ref\kuchar{K. Kuchar, in: General Relativity and Gravitation 1992,
R.J. Gleiser, C.N. Kozameh and O.M. Moreschi eds., IOP pub.,
Bristol, 1993.}; the form we obtain is perhaps sufficiently
attractive to deserve further investigation. It is also
encouraging that it agrees with what one would expect from
the work of
\ref\sam{J. Samuel,  \CQG{5} (1988) L123.}
\ref\CDJ{R. Capovilla, J. Dell and T. Jacobson, \PRL{63} (1989)
2325.}.

A more sophisticated treatment, along the lines of
\ref\brewin{L. Brewin, \CQG{5} (1988) 1193, {\bf 10} (1993) 947.}
is perhaps possible, but has not been attempted here.
%
\newsec{Generalities about Regge calculus.}
I shall here describe the main elements of Regge calculus
that are needed in the following,  in the
case of $3+1$ dimension, referring to ref.
\ref\WT{R.M. Williams , P.A. Tuckey,  \CQG{9} (1992) 1406.}
for a proper review and a complete bibliography.

4-space is divided in 4-simplices $\sigma^4$, each with a flat
metric
determined by the assigned lengths of its links $\sigma^1_\alpha ,
\ \alpha =1,...,10$. A Lorentz frame can therefore be attached to
each $\sigma^4$, through vierbeins $e^i_\mu (\sigma^4)$, constant
within $\sigma^4$, which associate to each link $\sigma^1_\alpha$ a
Lorentz vector $v^i(\sigma^1_\alpha |\sigma^4)=e^i_\mu (\sigma^n)
\Delta x^\mu_\alpha$.
Two neighbouring 4-simplices, say $\sigma^4_A,\ \sigma^4_B$ share a
tetrahedron $\sigma^3$; since the lengths of of all
$\sigma^1_\alpha\in\sigma^3$ must be the same in the two frames,
a Lorentz transformation will link them, such that:
\eqn\i{ v^i(\sigma^1_\alpha |\sigma^4_A)=\Lambda (A\,B)^i_{\ j}
v^j(\sigma^1_\alpha |\sigma^4_B),\quad
\forall\ \sigma^1_\alpha\in\sigma^3 =\sigma^4_A\cap \sigma^4_B
}
There are 3 independent such equations, which fix $\Lambda_{AB}$
uniquely\foot{
Assigning the action of an $O(n)\;$ (or $O(n-1,1)\;$)
transformation on $\ 1,\ 2,...,n-1\ $ independent  vectors fixes
$\ (n-1),\ (n-1)+(n-2),...,\ $ all its $n(n-1)/2\ $
parameters}.
This is the "metricity" condition; in the continuum the condition
that the torsion be zero fixes the Lorentz connection
$\omega^{ij}_\mu$ to be equal to the metric compatible
Levi--Civita connection $\Omega^{ij}_\mu$, here
distributional, with support on $\sigma^3$, and with
$\Lambda_{AB} = P\exp\int^B_A \Omega$.

Curvature is found going around a "bone" $\sigma^2=\sigma^4_1
\cap ...\cap\sigma^4_N$, where we find:
\eqn\ii{ R(\sigma^2|\sigma^4_1)=\Lambda (1N)...\Lambda (21)}
a Lorentz tansformation, which for consistency must be such that:
\eqn\iii{R(\sigma^2|\sigma^4_1)^i_{\ j}
v^j(\sigma^1_\alpha |\sigma^4_1) =
v^i(\sigma^1_\alpha |\sigma^4_1),\quad
\forall\ \sigma^1_\alpha\in\sigma^2  }
This gives us 2 independent equations, that by themselves
fix $R$ up to one parameter (the "deficit angle", which in fact
may or may not be an angle). Explicitely:
\eqn\iv{R = \exp (F),\qquad F^i_{\ j}=\alpha\,\epsilon^i_{\ jkl}
  v^k_1v^l_{2}}
In the Ashtekar approach the emphasis is on
the use and the transport of self-dual vectors, which are
projected from antisymmetric tensors by
\eqn\vi{ v^a := C^a_{ij}V^{ij} = - \half \epsilon_{abc}V^{bc}+i\,V^{0a}\ ;
\quad  a,b,...=1,2,3 }
and  transform  under the $(1,0)$ representation of the Lorentz
group:
\eqn\vii{\Lambda^i_{\ j}=\delta^i_{\ j}+\lambda^i_{\ j}+...\
\Longleftrightarrow
\ L_{ab}=\delta_{ab}+\epsilon_{acb}C^c_{ij}\lambda^{i j}+...
}
The $3\times 3$ matrices $L_{ab}$ are orthogonal and (in general)
complex; for an ordinary rotations they are real with
$L_{ab}=\Lambda_{ab}$ ; for a Lorentz boost
(a pure Lorentz transformation) they are hermitian.
Selfduality means that
$\half \epsilon_{ij}^{\ \ kl}C^a_{kl}=iC^a_{ij}$, therefore to
$\Lambda^i_{\ j}=\delta^i_{\ j}+\epsilon^i_{\ jkl}\Delta^{kl}+...$
corresponds
$L_{ab}=\delta_{ab}+2i\epsilon_{acb}C^c_{ij}\Delta^{i j}+...$.

The arguments above can be
repeated focusing on bones rather than links, associating to
a bone $\sigma^2$ with given $S^{\mu\nu}$ a 3--vector by
$v^a_{\rm  bone}=C^a_{ij} e^i_\mu e^j_\nu S^{\mu\nu}$, in each frame.
Notice that there are 10 links, but also 10 triangles for each
$\sigma^4$, so that assigning the areas of the triangles fixes
the metric just like assigning the lengths of the links
\ref\carloc{C. Rovelli, Pittsburgh and Trento preprint, 1993.}.

Going round a bone:
\eqn\viii{R=\exp (F)\qquad F_{ab}:=\epsilon_{acb}F^c=
2i\alpha\epsilon_{acb}v^c_{\rm  bone}\quad
{\rm real}\ \alpha
}
which follows because $R$ has to leave the bone invariant.
%
\newsec{3+1 separation.}

To separate space from time  an appropriate set of $\sigma^3$-s
has to be selected as a Cauchy surface of constant time.
Each inherits a Lorentz frame from (say) its future $\sigma^4$;
but while the lengths of the $\sigma^1_\alpha\in\sigma^3$ fix
the (flat) 3--metric, to determine the Lorentz vectors
$v^i_\alpha$, hence the $e^i_\mu ,\ i=0,1,2,3,\ \mu =1,2,3$
we need also informations about the movement of those links
in time.  However, since the lengths are positive,
the 3--metric $q_{\mu\nu}=\eta_{ij}e^i_\mu e^j_\nu$ must be
positive definite; therefore for each $\sigma^3$ there
is a Lorentz transformation $\Lambda$ such that
$(\Lambda e_\mu )^0 = -n_\mu$,
the unit normal the the 3--space hypersurface, i.e.
such that $(\Lambda v(\sigma^1|\sigma^3))^0= 0$. This choice of
frames we shall call "time gauge", and will be used a lot in
the following.

A Lorentz transformation links the frames of a pair
$\sigma^3_A,\ \sigma^3_B$ sharing a $\sigma^2$, the same that
links the corresponding $\sigma^4$-s. But if we only have access
to the spacelike links, we can only get 2 independent equations
from eq.\i , and the Lorentz transformation is fixed only
up to one parameter: $\Lambda (AB)$  depends on the
3--d Levi--Civita connection, but also on the extrinsic curvature
at  $\sigma^2$.
Suppose we choose the time gauge for both $\sigma^3_A,\ \sigma^3_B$,
and write  $\Lambda (AB)$ as a rotation times a boost. Since
$v^0_\alpha (A)=v^0_\alpha (B)=0$, the
boost must be {\it transversal} to all $v_\alpha (B) \in
\sigma^2$, which fixes it up
to a constant (related to the extrinsic curvature), and
the 3--rotation is uniquely determined. This transversality
of the Lorentz boost is an essential ingredient in G. 't Hooft's
treatment of the $2+1$ dimensions  theory \thooft .

In the continuum we have the analogous statement
that pulling back the 4--d Levi-Civita connection one has:
\eqn\v{q_\mu^{\;\nu}\Omega^{ab}_\nu = \3\Omega^{ab}_\mu\ ;
\quad q_\mu^{\;\nu}\Omega^{0a}_\nu = e^{a\nu}K_{\mu\nu}\qquad
a,b =1,2,3 \ ;\quad {\rm  (time\ gauge)}
}
Where $\3\Omega^{ab}_\mu$ is the 3--d Levi--Civita connection,
$K_{\mu\nu}=K_{\nu\mu}$ the extrinsic curvature.

Going round a $\sigma^1=\sigma_1^3\cap ...\cap\sigma_N^3$ we
get the curvature associated by eq.\ii\ to the corresponding
timelike bone: a Lorentz transformation, determined up to
3 parameters by the condition that it leaves invariant the
vector associated to $\sigma^1$.

The canonical formalism of A. Ashtekar \abhayb\  is obtained
separating time from space in the selfdual picture:
one defines "new variables" from the pull backs of the
connection and the vierbein forms by:
\eqn\ix{A^a_\mu := q_\mu^{\;\nu}C^a_{ij}\omega^{ij}_\mu\ ;\qquad
\tilde E^{a\mu} := -C^a_{ij} e^i_\nu e^j_\lambda
\tilde\epsilon^{\mu\nu\lambda}=-2iC^a_{ij}e\,n_\nu e^{i\nu}e^{j\mu} }
 where $e=\det (q_{\mu\nu})^{1/2}$.
In time gauge and for the Levi-Civita connection:
\eqn\x{A^a_\mu =-\half\epsilon_{abc}\3\Omega^{bc}_\mu +ie^{a\nu}
K_{\mu\nu}\ ;\qquad
 \tilde E^{a\mu} = e e^{a\mu}\qquad {\rm  (time\ gauge)}
}
The $ \tilde E^{a\mu}$ associate quite naturally (complex)
 3--vectors to surfaces in 3-space
\ref\carlob{C. Rovelli, \PR{ D47} (1993) 1703.}, by
\eqn\xi{S^a =  \tilde E^{a\mu}\epsi_{\mu\nu\lambda}S^{\nu\lambda}=
 -2\, C^a_{ij} e^i_\mu e^j_\nu S^{\mu\nu}=\ {\rm  (in\ time\ gauge)\ }
\epsilon_{abc}e^b_\mu e^c_\nu S^{\mu\nu}
}
For example for a triangle $(\Delta x_1,\;\Delta x_2,\;
 \Delta x_1-\Delta x_2)$,
$S^{\mu\nu}=\half\Delta x_1^{[\mu}\Delta x_2^{\nu ]}$,
so we find
$S^aS^a=\quart (\Delta x_1^2\Delta x_2^2 -(\Delta x_1\Delta x_2)^2)$
 = area square. We therefore have the necessary consistency condition:
\eqn\xii{S^a =  \tilde E^{a\mu}s_\mu\ \Rightarrow\ S^aS^a =
{\rm  real,\ positive}\quad \forall\; s_\mu {\rm \; real}
 }
a first set of "reality conditions", from which the existence of a
time gauge follows. In time gauge the $\tilde E^{a\mu}$-s, and
therefore any $S^a$, are real. In a different frame $S^a$ will
be complex, with (orthogonal) real and imaginary parts that carry
the informations about the surface and its orientation, and about
the components of its velocity perpendicular to this orientation.

In a discretized world, if tetrahedra $\sigma^3_A$ and
$\sigma^3_B$ share
a triangle to which $S^a_A,\ S^a_B$ are associated, one passes from
from one to the other by a Lorentz transformation:
\eqn\xiii{L(A\,B)= P\exp\int^B_A A\ :\qquad S_A^a=L(A\,B)_{ab}S_B^b
 }
 This metricity condition corresponds to the continuum
"Gauss law": $D_\mu\tilde E^{a\mu}=0$. However it is not really
enough, because it fixes $L_{AB}$
only up to one {\it complex} parameter. In the continuum the
missing information is supplied  by a second set of
reality conditions:
\eqn\xiv{\tilde P^{\mu\nu}:=i\epsilon_{abc}\tilde E^{c\rho}
\big (\tilde E^{a\mu}D_\rho\tilde E^{b\nu}+
\tilde E^{a\nu}D_\rho\tilde E^{b\mu}\big )=
\ {\rm  real}
}
If the connection is the pull back of the selfdual part of the
4-d Levi--Civita connection, one finds
$\tilde P^{\mu\nu} = 2e^3(K^{\mu\nu}-q^{\mu\nu}K)$.

To be definite, suppose that the tetrahedra
$\sigma^3_B,\;\sigma^3_A,\;\sigma^3_C$,
all sharing a common link $\sigma^1$, are
separated by the triangles $\sigma^2_1,\,\sigma^2_2,\,\sigma^2_3$,
so that $\sigma^3_A$ has sides $\sigma^2_1,\;\sigma^2_2$ etc.,
and $S_I^{\mu\nu} = \half \Delta x^{[\mu}\Delta x_I^{\nu ]}$,
$S_{I\mu}:=\half \epsi_{\mu\nu\lambda}S^{\nu\lambda}_I$,
with $I=1,\,2,\,3$.

At the boundary between  $\sigma^3_B$ and $\sigma^3_A$,
the covariant derivative of $\tilde E^{a\mu}$ is
non zero only in the direction normal to $\sigma^2_1$, so a
plausible translation of eq. \xiv\ is:
\eqn\xv{i\,\epsilon_{abc}\tilde E^{c\lambda}_AS_{1\lambda}\big (
 \tilde E^{a\mu}_AL_{bd}\tilde E^{d\nu}_B
 \,+\,\tilde E^{a\nu}_AL_{bd}\tilde E^{d\mu}_B \big ) =
 {\rm  real}
 }
To interpret this condition, let us associate a
(densitized) 3-vector to a link by:
 \eqn\xvi{\tilde l^a := \half\epsilon_{abc}\tilde E^{b\mu} \tilde E^{c\nu}
 \epsi_{\mu\nu\lambda}\Delta x^\lambda
 =-2ieC^a_{ij} e^i_\mu e^j_\nu n^\mu \Delta x^\nu =\ {\rm  (in\ time\ gauge)\ }
 e e^a_\mu\Delta x^\mu
 }
Notice that if (in time gauge) we organize the frames in two
neighbouring tetrahedra so that the Lorentz transformation from
one to the other is just a boost transversal to the common triangle,
the $\tilde l^a$ associated to
the sides of the triangle will {\it not} be invariant.

{}From the definitions one can derive the identities:
\eqn\xvii{S^a_{IA}\tilde l^b_A =S^a_{IA}\tilde l^b_{IA}=0\ ;\quad
\epsilon_{abc}S^b_{1A}S^c_{2A}=\quart\;
     \tilde l^a_A\;\epsi_{\lambda\mu\nu}
\Delta x^\lambda\Delta x_1^\mu\Delta x_2^\nu
\ ;\quad  \epsilon_{abc}\tilde l^b_A\tilde l^c_{IA}=e^2_A S^a_{IA}
 }
(for $I=1,\,2$), that we use to prove that:
\eqn\xviii{\eqalign{
\epsilon_{abc}\tilde l^b_A L_{cd}\tilde l^d_B =&
k\epsilon_{abc}\epsilon_{bde}\epsilon_{cfg}S^{d}_{1A}S^{e}_{2A}
L_{ff'}S^{f'}_{1B}L_{gg'}S^{g'}_{2B}=\cr
&= kS_{1\lambda}\tilde E^{a\lambda}_A\;
 \epsilon_{bcd}\tilde E^{b\rho}_A
\tilde E^{c\mu}_AL_{de}\tilde E^{e\nu}_B
S_{1\rho}S_{2\mu}S_{2\nu} \cr}
}
with $k$   real; by  eq.\xv\ then:
\eqn\xix{
 \epsilon_{abc}\tilde l^b_A\, L_{cd}\tilde l^d_B = i\, h\,S^a_{1A}
 }
with $h$ real. We may take this relation, which must hold
for any of the sides of the triangle we cross going from one
tethrahedron to another, as "reality condition" for $L_{AB}$,
and we shall presently see that together with eq.\xiii , it
gives us an adequate characterization of $L_{AB}$.

It follows from eqs.\xiii\ and \xix\ that
that the action of $L_{AB}$ on $\tilde l_b$ must be of the form:
\eqn\xx{L_{ab}\tilde l^b_B =a\,\tilde l^a_A+i\,b\,\epsilon_{abc}
\tilde l^b_A S^c_{1A}
}
with $a$ and $b$ real constants. Furthermore, since
$\tilde l^a_B,\ S^a_{1B},\ \epsilon_{abc}\tilde l^b_BS^c_{1B}$
are orthogonal, we can align the orthonormal basis in $B$ along
their directions, do the same in $A$, then use eqs.\xiii ,\ \xix\
to write the transformation in the form:
\eqn\xxi{L_{AB}=\left (\matrix{\cosh\eta_1&0&-i\sinh\eta_1\cr
 0 & 1 & 0 \cr i\sinh\eta_1&0&\cosh\eta_1}\right )
 = e^{-i\eta_1A_2}\ ;\quad (A_2)_{ab}=\epsilon_{a2b}
}
Let us complete the picture, going all the way round
the link $\sigma^1$. After the transformation \xxi\
we  change again the basis aligning it with
$\tilde l^a_A,\ S^a_{2A},\
\epsilon_{abc}\tilde l^b_AS^c_{2A}$ by a rotation
\eqn\xxii{R_{12}= \left (\matrix{1&0&0\cr
0&\cos\theta_A & \sin\theta_A \cr
0&-\sin\theta_A &\cos\theta_A\cr}\right ) =
e^{-\theta_A A_1}\ ;\quad (A_1)_{ab}=\epsilon_{a1b}
}
In this way we go from $B$ to $A$, to $C$ and back to $B$
round the link, rotating vector components in $B$ by:
\eqn\xxiii{R =   e^{-\theta_B A_1}e^{-i\eta_3 A_2}e^{-\theta_C A_1}
e^{-i\eta_2 A_2}e^{-\theta_A A_1}
e^{-i\eta_1 A_2}= e^{\alpha A_1+i\beta A_2+i\gamma A_3}:= e^F
}
More explicitely, we find that $F^c=\half\epsilon_{acb}F_{ab}$,
in the frame of $B$, must have the form
\eqn\xxiv{F^c_B=a\,\tilde l^c_B+ib\, S^c_{1B}
+ic\,\epsilon_{cde}\tilde l^d_BS^e_{1B}
}
with $a,\;b,\;c$ real constants.
This structure of $F$ agrees with what we should have expected from
eq.\viii : viewed in 4-space, we have gone round a bone with sides
$\Delta x^\mu$ and a time--like link that we may represent by
$(Mn^\mu +M^\mu )$, hence from eqs.\viii , \xvi\
$F^c$ must be of the form :
\eqn\xxv{F^c= 2i\alpha C^c_{ij}e^i_\mu e^j_\nu
(Mn^\mu +M^\mu )\Delta x^\nu =-\alpha M \tilde l^c_B + i\,S'^c
\ ,\quad \tilde l^a_B  S'^a =0
}
{}From this expression we also see that, since one link has to be
time--like:
 \eqn\xxvi{F^aF^a>0}
This concludes our "kinematic" analysis of Regge calculus in the
selfdual picture.

\newsec{The constraints.}
What makes gravity in $3+1$ dimensions so much more
interesting than in $2+1$ dimensions is that $F^a$ does not
vanish, but has to satisfy first class constraints \abhayb .
To be able to do "dynamics" we have to express these
constraints in the Regge case.

To get an idea of what might be their form, we shall consider the
situation for a smooth manifold with weak field. In this case
we succeed in expressing the constraints in a form
which uses the notions introduced in the previous section; this
might be a useful starting point.

To begin with, consider a triangle $(1,\,2,\,3)$, and set
$x^\mu_{\alpha\beta}:=\half
(x_\alpha +x_\beta )^\mu$,  $\Delta^\mu_{\alpha\beta}:=
(x_\alpha -x_\beta )^\mu$; then we find:
\eqn\xxx {
x^\mu_{12}\Delta^\nu_{21}+ x^\mu_{23}\Delta^\nu_{32}+
 x^\mu_{31}\Delta^\nu_{13}=
x_1^{[\mu}x_2^{\nu]}+ x_2^{[\mu}x_3^{\nu]}+x_3^{[\mu}x_1^{\nu]} =
2S^{\mu\nu}
}
from which for example we may derive, for small $A$, the
well known equation:
\eqn\xxxi{
\exp (A(x_{12})_\nu\Delta^\nu_{21})
\exp (A(x_{23})_\nu\Delta^\nu_{32})
 \exp (A(x_{31})_\nu\Delta^\nu_{13})=
 1+F_{\mu\nu}S^{\mu\nu}+\ldots
}
with:
\eqn\xxxii{A_{\mu}^{ab}=\epsilon_{acb}A^c_\mu\ ;\quad
F_{\mu\nu}^{ab}=\epsilon_{acb}F^c_{\mu\nu}
= \epsilon_{acb}(\partial_\mu A^c_\nu -\partial_\nu A^c_\mu
+\epsilon_{cde}A^d_\mu A^e_\nu )
}
and $F_{\mu\nu}$ calculated somewhere in the middle of the triangle.

In a similar way, for a tetrahedron $(1,2,3,4)$, setting:
\eqn\xxxiii{\eqalign{
&x^\mu_{\alpha\beta\gamma}:=
{\textstyle{1\over 3}}(x_\alpha^\mu +x_\beta^\mu +x_\gamma^\mu )
\ ;\quad
S^{\mu\nu}_{\alpha\beta\gamma}:= \half (
x_\alpha^{[\mu}x_\beta^{\nu]}+
x_\beta^{[\mu}x_\gamma^{\nu]}+x_\gamma^{[\mu}x_\alpha^{\nu]})\cr
&\V  :={\textstyle{1\over 6}}\epsi_{\mu\nu\lambda}
(x_1^\mu x_2^\nu x_3^\lambda +x_2^\mu x_1^\nu x_4^\lambda +
x_3^\mu x_4^\nu x_1^\lambda +x_4^\mu x_3^\nu x_2^\lambda  )\cr
&\tilde l^a_I :=\half\epsilon_{abc}\tilde E^{b\nu}
\tilde E^{c\lambda} \epsi_{\mu\nu\lambda}x_I^\mu
\qquad {\rm  for}\quad I:= (123),\ (214),\ (341),\ (432)  \cr}
}
one finds:
\eqn\xxxiv {
\sum_Ix_I^\mu\, S_I^{\nu\lambda} =
\half \tilde\epsilon^{\mu\nu\lambda}\V
}
To use this identity in a Regge-like situation,
suppose we take the origine where four links meet, in such
a way that the $x_I^\mu $ are  links of a Regge lattice;
then the tetrahedron itself belongs to
the {\it dual} lattice, the triangles $S_I$ giving paths that
encircle the links. With this picture in mind we can derive:
\eqn\xxxv {\eqalign{
\sum_I\epsilon_{acb}F^c_{\nu\lambda}S_I^{\nu\lambda}
\tilde l_I^b &=
\half \epsilon_{acb}F^c_{\mu\nu}\epsilon_{bde}\tilde E^{d\rho}
\tilde E^{e\sigma} \epsi_{\rho\sigma\lambda}
\tilde\epsilon^{\lambda\mu\nu}\V = 2\,\tilde E^{a\mu}
\tilde E^{b\nu}F^b_{\mu\nu}\V  \cr
\sum_IF^a_{\mu\nu}S_I^{\mu\nu} \tilde l_I^a  &=
\half F^a_{\mu\nu}\epsilon_{abc}\tilde E^{b\rho}
\tilde E^{c\sigma} \epsi_{\rho\sigma\lambda}
\tilde\epsilon^{\lambda\mu\nu}\V = \epsilon_{abc}\tilde E^{a\mu}
\tilde E^{b\nu}F^c_{\mu\nu}\V \cr}
}
The r.h.s. of these identities has the form of the vector and
scalar constraints in the Ashtekar formulation, and should
therefore be set to zero. These identities then provide
what may be an appealing interpretation of the constraints,
very similar to the interpretation given by K. Kuchar\kuchar
to the properties of the curvature in triad dynamics. Notice
that the real-imaginary structure of $F^c$ we found in the
previous section makes the first line of eqs.\xxxv\ pure imaginary,
the second real, as they should be \me .

However to apply these identities to Regge
calculus one should assume that the  $F^c$--s associated to the
different links that meet at some vertex can be obtained from
a single $F^c_{\mu\nu}$, contracted with the different surface
elements, which seems unlikely. It would appear more likely,
if anything, to conjecture that a single $F^c_{\mu\nu}$ is
associated to each $\sigma^3$, and that the different $F^c_l$--s
one finds going round its six links $l=(\alpha ,\beta )$ can
be obtained contracting
this $F^c_{\mu\nu}$ with the surfaces of six polygons belonging
to the dual lattice.

Tentatively, we can try and simulate the actual
dual lattice taking an origine $0$ at the barycenter of
$\sigma^3$,  and fixing
points $\s 1,\; \s 2,\; \s 3,\; \s 4,\; $, with coordinates $\s x_1^\mu
=k(x_2^\mu +x_3^\mu +x_4^\mu )$, ...,  $k$ large enough so
that the triangles $(0,\, \s 1,\,\s 2)$, ... encircle the links
$(3,\,4)$, ...;  define then:
\eqn\xxxvii{\eqalign{
&\Delta_{\alpha\beta}^\mu :=x^\mu_\alpha -x^\mu_\beta\ ;\quad
 \s S^{\mu\nu}_{\alpha\beta}:=\sum_{\gamma\delta}
\quart\varepsilon_{\alpha\beta\gamma\delta}
 {\s x^{[\mu}_\gamma}{\s x^{\nu ]}_\delta}\ ,
\quad\varepsilon_{\alpha\beta\gamma\delta}=\pm 1\;,\
\varepsilon_{1234}=+1\cr
&\tilde l^a_l :=\half\epsilon_{abc}\tilde E^{b\nu}
\tilde E^{c\lambda} \epsi_{\mu\nu\lambda}\Delta_l^\mu
\quad {\rm  for}\quad l:=(1,2),\;(2,3),\;(3,4),\;(1,4),\;(4,2),\;(3,1)
\cr}
}
What we are supposing is that:
 \eqn\xxxviii {
F^c_l=F^c_{\mu\nu} \s S_l^{\mu\nu}
}
To write the constraints we need another elementary  geometric
identity, similar to eq.\xxxiv , that again follows
from the definitions:
\eqn\xxxix{
\sum_l \Delta_l^\mu \s S_l^{\nu\lambda}
 = {3\over 2}k^2\V \tilde\epsilon^{\mu\nu\lambda}
}
The Ashtekar constraint would then be, for each tetrahedron:
\eqn\xl{\eqalign{
\sum_l \epsilon_{acb}F^c_l \tilde l_l^b  &=
{3\over 4}k^2\epsilon_{acb}F^c_{\mu\nu}\epsilon_{bde}\tilde E^{d\rho}
\tilde E^{e\sigma}\,\epsi_{\rho\sigma\lambda}
\tilde\epsilon^{\lambda\mu\nu}\V = 3k^2\,\tilde E^{a\mu}
\tilde E^{b\nu}F^b_{\mu\nu}\V =0 \cr
\sum_l F^a_l \tilde l_l^a   &=
{3\over 4}k^2 F^a_{\mu\nu}\epsilon_{abc}\tilde E^{b\rho}
\tilde E^{c\sigma} \epsi_{\rho\sigma\lambda}
\tilde\epsilon^{\lambda\mu\nu}\V =
{3\over 2}k^2\epsilon_{abc}\tilde E^{a\mu}
\tilde E^{b\nu}F^c_{\mu\nu}\V =0  \cr}
}
To test whether this form of the constraints is correct would
presumably require a careful derivation of the $3+1$ action from
the Regge action, along the lines of \brewin ; and to see whether
it is a useful form one would have to develop a proper canonical
formalism, and test for the closure of the constraints under
Poisson brackets, a feat so far achieved only in $2+1$ dimensions
\wael .

Notice that the idea that there should be an $F^a_{\mu\nu}$ for each
tetrahedron is consistent with the solution of the constraints
proposed in \sam , and generalized in \CDJ  to
\eqn\xlii{ F^a_{\mu\nu}=\Phi_{ab}\;
 \epsi_{\mu\nu\lambda}\tilde E^{b\lambda}
}
with $\Phi_{ab}$ a traceless symmetric tensor, since in Regge
calculus there is certainly an $\tilde E^{a\mu}$ for each
tetrahedron; it is unfortunate that I cannot suggest a geometric
interpretation for $\Phi_{ab}$.

I would like to thank Prof. Abhay Ashtekar for the warm
hospitality at the University of Syracuse, Prof. V. Radhakrishnan
for the hospitality at the Raman Research Institute in
Bangalore, and Mauro Carfora, Maurizio Martellini, Annalisa
Marzuoli and Joseph Samuel for the encouragement that has made
this work possible.

\listrefs
\bye